\documentclass[10pt,journal,compsoc]{IEEEtran}
\usepackage{quoting}
\usepackage[center]{caption}
\usepackage{booktabs}
\usepackage[table,xcdraw]{xcolor}
\usepackage{caption}
\usepackage{subcaption}
\usepackage{soul} 
\usepackage[skins]{tcolorbox}
\usepackage[]{graphicx}

\newtcolorbox{myframe}[2][]{%
  enhanced,colback=white,colframe=black,coltitle=black,
  sharp corners,boxrule=0.4pt,
  fonttitle=\itshape,
  attach boxed title to top left={yshift=-0.3\baselineskip-0.4pt,xshift=2mm},
  boxed title style={tile,size=minimal,left=0.5mm,right=0.5mm,
    colback=white,before upper=\strut},
  title=#2,#1
}

\ifCLASSOPTIONcompsoc
  \usepackage[nocompress]{cite}
\else
  \usepackage{cite}
\fi
\begin{document}
%
\title{What Makes a Great Software Quality Assurance Engineer?}
%
%
%
%

\author{Roselane~Silva Farias, 
        	Iftekhar~Ahmed, 
        and~Eduardo Santana de Almeida,~\IEEEmembership{Senior Member,~IEEE}
}

\IEEEtitleabstractindextext{%
\begin{abstract}
Software Quality Assurance (SQA) Engineers are responsible for assessing a product during every phase of the software development process to ensure that the outcomes of each phase and the final product possess the desired qualities. In general, a great SQA engineer needs to have a different set of abilities from development engineers to effectively oversee the entire product development process from beginning to end. Recent empirical studies identified important attributes of software engineers and managers, but the quality assurance role is overlooked. As software quality aspects have become more of a priority in the life cycle of software development,
employers seek professionals that best suit the company's objectives and new graduates desire to make a valuable contribution through their job as an SQA engineer, but what makes them great? We addressed this knowledge gap by conducting 25 semi-structured interviews and 363 survey respondents with software quality assurance engineers from different companies around the world. We use the data collected from these activities to derive a comprehensive set of attributes that are considered important. As a result of the interviews, twenty-five attributes were identified and grouped into five main categories: personal, social, technical, management, and decision-making attributes. Through a rating survey, we confirmed that the distinguishing characteristics of great SQA engineers are curiosity, the ability to communicate effectively, and critical thinking skills. This work will guide further studies with SQA practitioners, by considering contextual factors and providing some implications for research and practice.
\end{abstract}

}

\maketitle

\IEEEdisplaynontitleabstractindextext

%
\IEEEpeerreviewmaketitle

\IEEEraisesectionheading{\section{Introduction}\label{intro}}

\IEEEPARstart{S}oftware Quality Assurance is an important and integral part of today’s software industry. 
Throughout the evolution of software development methods, several factors have contributed to the importance of the Software Quality Assurance (SQA) role.
The increased coupling of the internet to many software applications is one, slightly dated, but important factor as it requires more attention to analyzing products for security and performance attributes. A second factor, the increasing demand for more frequent software releases, increased the need to optimize the validation and verification processes so the product is delivered faster and with increased quality \cite{review:2018}.

Software testing has evolved from being a means of finding and reporting bugs to a key ingredient in the broader software quality assurance process. Agile and Scrum process models are major contributors to changing the way testing is integrated throughout the entire development cycle, increasing the importance of testing in the software development methodology \cite{8048665}. 
As software testing evolves, SQA engineers are expected to address a broader array of attributes.

Some evidence suggests that attributes of SQA engineers should rely on their ability to communicate and collaborate among other SQA members and development teams \cite{review:2018}. This is mainly observed with scrum-type processes, where there is an agile team working to achieve a common goal.

In addition, an SQA engineer should likely know the ``big picture", i.e., have full visibility into product requirements and customer needs \cite{review:2018} to analyze what should be covered during testing. Finally, an SQA engineer will most likely seek ways to optimize testing efforts, which requires other attributes, such as technical skills \cite{santamaria:2007}. Although the literature has indicated some attributes an SQA engineer should have (\cite{santamaria:2007,review:2018}), no studies validate these attributes from the perspective of what SQA engineers think the attributes of a great SQA engineer are and the importance of these attributes.
 
There are empirical studies that have considered attributes for other software engineering roles. Li et al. conducted two empirical studies to identify a set of attributes that expert Software Engineers viewed as important to becoming a great software engineer \cite{Li:2015, Li:2020}. 
Kalliamvakou et al.\cite{Kalliamvakou:2019} interviewed managers of Software Engineers to uncover attributes crucial for Software Managers.
Recently, Dias et al. \cite{dias:2021} conducted an empirical study with experienced open-source software maintainers to identify the attributes that describe great maintainers. 

In the context of software testing, some studies address testing techniques \cite{7814898}, automation tools \cite{mariani:2017}, testing practices \cite{8048665}, and competencies for software testing roles \cite{kassab:2021} among others. However, to the best of our knowledge, there is a lack of studies that characterize the integrated SQA role and the important attributes that this professional should have. Therefore, we sought to mitigate the lack of characterization of a great SQA engineer by investigating the following research questions: 

\begin{itemize}
    \item RQ1 - What are the attributes of great SQA engineers? Why are these attributes viewed as important?

    \item RQ2 - How does the SQA community perceive the importance of these attributes?
    
    \item {RQ3 - How do the attributes of a great SQA engineer differ from other software development roles?}    
    
\end{itemize}

To answer these research questions, we conducted 25 semi-structured interviews with SQA engineers from different companies around the world. The main contribution of the interviews was an initial understanding of the software quality assurance engineer role and a list of 25 attributes of great SQA engineers split into five different categories: \textit{personal, social, technical, management, and decision-making} attributes. To further understand the importance of each identified attribute, we performed a large-scale survey with 363 expert SQAs. We found that the top five distinguishing characteristics of great SQAs are curiosity, the ability to communicate effectively, critical thinking skills, learning continuously, and technical expertise. 

This paper constitutes the first empirical study with SQA engineers on characterizing the integrated SQA role and identifying the attributes that great SQAs should possess and the importance of these attributes.

It makes the following contributions:

\begin{itemize}
    \item A mixed method approach that investigates the software quality assurance role and the attributes of great quality assurance engineers.
    
    \item A set of twenty-five attributes 
    that define the roles and responsibilities of SQA engineers and offer implications for practitioners, researchers, and educators.

    \item A collection of all our research materials on a project website for replication and reproducible research purposes, including our interview data (prompts, transcriptions, and codebook) and the survey instrument. 
    \footnote{https://github.com/great-SQA/empirical-study-great-SQA}
\end{itemize}

The remainder of this paper is structured as follows. Section \ref{relatedwork} discusses our work in the context of similar empirical studies about other software development roles. Section \ref{methodology} outlines our research methodology. The set of attributes identified from the interviews (RQ1), the importance of the identified attributes (RQ2) with analysis by two contextual factors, and a comparison with previous studies (RQ3) are discussed in Section \ref{results}. Section 
\ref{discussion} provides a more detailed discussion and some implications for practitioners, researchers, and educators. Section \ref{threats} presents the key limitations and how we mitigated them. Finally, Section \ref{conclusion} presents the conclusions and future work.                                  
\section{Related Work}
\label{relatedwork}

\begin{table*}[!t]
\caption{Professional and demographic information of the participants. Their position at work, how many employees report to them (in case she/he is in a leadership role), experience in years as an SQA professional, and the country where the employee currently lives and where the company is located.}
\centering
\label{tablebig}
\begin{tabular}{p{0.15\linewidth}p{0.25\linewidth}p{0.15\linewidth}p{0.15\linewidth}p{0.15\linewidth}}
\hline
\textbf{P(n)} & \multicolumn{1}{l}{\textbf{Job Title}} &
\textbf{\begin{tabular}[c]{@{}c@{}}\# Direct \\reports\end{tabular}}
& \textbf{\begin{tabular}[c]{@{}c@{}}Experience \\ (years)\end{tabular}} & \textbf{\begin{tabular}[l]{@{}c@{}}Country \\ (employee/company)\end{tabular}} 
\\
\hline
P1 & SQA Engineer III &	0	&3-5 &	Brazil/USA\\
P2 & Senior SQA Engineer IV	&1&	5-10 &	Brazil/USA\\
P3 & Senior SQA Engineer IV	&0&	5-10 &	Brazil/USA\\
P4 & SQA Team Lead	&7&	5-10 &	Brazil\\
P5 & Senior SQA Engineer IV	&0&	5-10 &	Brazil/USA\\
P6 & SQA Engineer III	&0&	3-5 &	Brazil/USA\\
P7 & SQA Team Lead	&7&	5-10 &	USA\\
P8 & SQA Team Lead	&1&	3-5 &	Brazil/USA\\
P9 & Senior SQA Engineer IV&	0&5-10 &	Brazil/USA\\
P10 & SQA Specialist	&0&	10+ &	Brazil/USA\\
P11 & SQA Engineer III	&0&	10+ &	Brazil/USA\\
P12 & Senior SQA Engineer IV	&0&	5-10 &	Brazil/USA\\
P13 & Senior SQA Engineer IV&	0&	1-2 &	Brazil/USA\\
P14 & SQA Team Lead&	1&	5-10 &	Portugal\\
P15 & Junior SQA Engineer (I-II)	&0&	1-2 &	Brazil\\
P16 & Senior SQA Engineer IV	&0	&5-10 &	Brazil\\
P17 & SQA Team Lead	&3&	5-10 &	Brazil\\
P18 & Junior SQA Engineer (I-II)&	3&	3-5 &	Brazil\\
P19 & SQA Team Lead&	19	&3-5 &	India\\
P20 & SQA Specialist &	0&	10+ &	Brazil\\
P21 & SQA Team Lead &	4&	10+ &	Brazil\\
P22 & SQA Team Lead &	7&	10+ &	Brazil\\
P23 & SQA Engineer III &	0&	10+ &	Brazil\\
P24 & SQA Team Lead&	5&	10+ &	Canada\\
P25 & SQA Specialist &	0&	10+ &	USA\\
\hline
\end{tabular}
\end{table*}

There are four recent studies related to this work. The first one investigated what makes a great software engineer \cite{Li:2015}. The authors identified a set of attributes considered important for the engineering of software. They performed 59 semi-structured interviews with employees at Microsoft and organized 54 attributes into internal attributes of the engineer’s personality and ability to make effective decisions. For instance, they found that decision-making is a key part of software engineering and made some recommendations to both academia and industry.

In a later study, these researchers \cite{Li:2020} conducted a mixed-method approach with Software Engineers at Microsoft. They surveyed 1,926 expert engineers to assess the importance of the set of attributes of great engineers identified in the previous study. Also, some contextual factors were considered in this work, such as the amount of experience, gender, country, computer science educational background, and type of software. The top five attributes of great engineers are writing good code, adjusting behaviors to account for future value and costs, practicing informed decision-making, avoiding making others’ jobs harder, and learning continuously.

Another software development role investigated was the software manager. Kalliamvakou et al. \cite{Kalliamvakou:2019} interviewed 37 and surveyed 563 engineers and software managers to investigate what manager attributes developers and engineering managers consider important and why. One of the contributions was a conceptual framework of fifteen manager attributes that characterize great engineering managers. Some examples include: motivating the engineers, mediating communication, and being technical.

Dias et al. performed a mixed-method approach with 33 semi-structured interviews and a rating survey with 90 open-source software maintainers to help maintainers excel in their careers \cite{dias:2021}. The authors created a conceptual framework with 22 attributes that explain how these attributes are connected. For instance, the top 5 attributes of a great OSS maintainer were communication, quality assurance, community building, empathy, and vision. 

Overall, the previous studies are prescriptive, offering recommendations and providing some insights into why these attributes are viewed as important. Our study follows a similar methodology but looks at the software quality assurance perspective \cite{8048665}.

\section{Methodology}
\label{methodology}
\noindent
Our research methodology consisted of interviews with software quality assurance (SQA) engineers to identify the attributes they perceive as important attributes of a great SQA engineer. 

\subsection{The Software Quality Assurance Engineer position}
Software Testing is one of the knowledge areas classified by the Software Engineering Body of Knowledge
(SWEBOK) \cite{SWEBOK:2014} as the dynamic verification that a program provides expected behaviors on a finite set of test cases.

SQA can be interpreted in different ways and
words. In this study, we will use the term \textit{Software Quality Assurance (SQA) engineer} where its main responsibility is to attain better quality in software products. 
  
Ron Patton\cite[p.~333]{ron:2005} cited the following about an SQA professional: 

\begin{quotation}
``\textit{A Software quality assurance person's main responsibility is to examine and measure the current development process and find ways to improve it with a goal of preventing bugs from ever occurring}". 
\end{quotation}
 
\subsection{Interviews}
Aiming to answer RQ1 and RQ2, we analyzed almost 100 attributes mentioned during 25 interviews. 

We conducted semi-structured interviews \cite{brereton:2008}, which is a flexible type of interview, allowing the interviewer to add or remove questions as they see fit, during the flow of the interview \cite{seaman:1999}. 

Performing a pilot study was very important to assess different aspects of the interview protocol. During the pilot, we noticed that participants often referred to someone they have worked with in the past, so we included one question for them to describe why they think that this person was great at their job. It also gave us a better idea of the timing of the interview and we were able to prepare extra questions to use in the interview if we had time.

\textbf{1) Participants:} We used a convenience sampling approach to recruit the participants. During the interviews, we also used a snowballing approach, asking the participants to put us in touch with other SQA engineers who would apply. 
Another approach we used to invite participants to the interview was through the LinkedIn platform\footnote{http://www.linkedin.com}.
Table \ref{tablebig} shows the job title, experience, and demographic information of the 25 interviewees. For those SQA engineers who have a leadership position, we asked how many direct reports they have. We also captured both the country they live in and where their companies are located because some of them work remotely.  

\textbf{2) Before the interview:} 
We asked the participants to fill out a short questionnaire before the interview. It had a consent term explaining the objective of the study, what the research participants could benefit from it, and a statement that the participant could give up participating in the study anytime without any problem. Although the interview did not collect any personal data, only opinions on the subject in question, the consent form also emphasized that the identity and personal information of the participants are preserved and would be treated as strictly confidential. 

Once the respondent checked the option to agree with participation in the study, we collected some background and demographic information.
We followed some instructions suggested by two related work \cite{Kalliamvakou:2019, dias:2021} to better guide the interviews. We asked the participants to list, in order of priority, at least five attributes they consider the most important for a great SQA engineer. This way, we avoided participants rushing their answers and required them to provide focused responses during the interview since they had time to send them at their leisure. 

\textbf{3) Interview protocol:}
The interview was divided into three parts: we collected information about their experience, discussed the attributes they submitted in the form, and then had a deep discussion about someone they thought was a great SQA engineer. We started the interview by describing the study, asking permission to record the meeting, and asking the following questions about their job:

\begin{itemize}
   \item What companies and what software testing roles, have you worked at?
   \item With Which testing types have you worked? We selected a list of testing types from
   \cite[p.~16]{rios:2013}. Some examples used in the interview were: acceptance testing, end-to-end testing, functional testing, integration testing, and regression testing.
   \item Do you work with manual test cases, automated test cases, or both?
   \item Do you work with any agile methodology?
   \item How many SQA engineers do you have on your team?
   \item Tell me more about your daily tasks.
\end{itemize}  

Then, for each attribute sent before the interview, we discussed the following:
\begin{itemize}
   \item Why do you think this attribute is important?
   \item Can you think of any personal experience you or someone on your team related to this attribute?
\end{itemize}    

The questions about each attribute varied according to their experiences. Finally, in the last part of the interview, we asked the participants to take a minute to think back to someone they worked and/or worked with in the past that they thought was a great SQA engineer. \begin{itemize}
   \item What are some attributes that made this person `great' in their mind?
   \item Why do they think this person was great at his job?
\end{itemize}    

We closed the interview by thanking the participants and asking interviewees whether they had final comments to add and whether they could suggest other SQA professionals that could be interviewed.

\textbf{4) Interview analysis:} 
The interviews were conducted remotely using the online channel Google Meet. On average, the interviews lasted 40 minutes (min: 35, max: 55). All the interviews were recorded with the consent of the interviewees, then the audio files were transcribed almost literally and integrally using two online transcription products: Otter.ai - for interviews in English\footnote{Otter.ai is available at https://otter.ai} and Trint - for interviews in Portuguese\footnote{Trint is available at https://trint.com}.

We analyzed 17 hours of interviews by performing open coding, which is a technique for performing qualitative analysis in grounded theory \cite{corbin:1998}. The qualitative data analysis software used to facilitate the coding process was \textit{QDA Miner}\footnote{QDA Miner is available at https://provalisresearch.com}. Coding is an interactive process, where the researcher reads through each transcription as many times are needed, creating and assigning codes. These codes are later grouped into categories that can be added, edited, deleted, refined, or even merged during the analysis. The process continues until the saturation point is achieved, e.g., no more codes and categories are needed when analyzing a new text \cite{farias:2019}.

As Strauss and Corbin suggest at least 10 interviews with detailed coding are necessary for a qualitative analysis \cite[p.~281]{corbin:1998}, we believe that conducting 25 interviews is a considerable sample size for this study.

We started the analysis process by reading the interview transcripts and assigning codes to pieces of text that the authors considered important to answer the research questions. We analyzed every attribute listed in the transcription and the sentence tied to it and added labels (codes) to these statements. Then, we grouped these codes of attributes into high-level categories. When we found the same attribute described differently, we grouped those under the same code. Finally, we categorized the attributes in higher-level clusters, based on similarities between the meanings of the codes. 

After several rounds of manual analysis and discussion among three authors, we reached a consensus and grouped all the attributes into five categories: decision-making, management, social, technical, and personal attributes. 

\subsection{Survey}
The final phase of our study was aimed at
validating the results that emerged from the interviews. For the design of the surveys, we followed Kitchenham and Pfleeger’s guidelines for personal opinion surveys \cite{Kitchenham:2008}. The survey was anonymous to increase response rates and the survey instrument is available on our project website. 

\textbf{1) Design of the Survey Questions:}
We designed a 15-minute survey to uncover the attributes of great software quality assurance (SQA) engineers. We used the results from the interviews with respondents to rate the level of agreement about how important each attribute is for distinguishing a great SQA engineer from a good one.

We started our survey by explaining the goal of the study and getting the consent form for volunteer participation. Then there was a section for demographic information, and the next section was intended to collect data about the SQA activities performed by them and the types of testing they usually work with. The participants were asked to rate the importance of the attributes identified in the interviews presented per category. In total, each participant rated 25 attributes, and the statements were on a 5-point Likert scale, from Very Unimportant to Very Important. 

\textbf{2) Participants:}
We recruited SQA professionals to participate by searching on LinkedIn, which contains resumes of professionals around the world. We mainly searched for LinkedIn members who were part of the “Software Test” or "QA" groups, which included more than 65 thousand members at the time of the study. The survey collected responses from February 15th to March 3rd, 2022. After a few weeks, we closed the survey with a total of 363 answers. 

The participants identified themselves as the following: 62.8\% male, 36.6\% female, 0.3\% non-binary, and 0.3\% preferred to not answer. 
Of the males, 69\% of them are in a management position while only 31\% of the female are in a management position. Next, we detail more demographic information about the participants.

\begin{itemize}
 
    \item \textit{Nationalities.} There were respondents across 18 nationalities spread over all continents except for Africa, while 203 answers came from North America (55.92\%).
    
    \item \textit{Educational qualification.} 
6\% of the participants only have high School, 5\% took technical school, 55\% have bachelor's degrees, and 34\% have Master's, Ph.D., or M.D.

\item \textit{Experience in the field of software quality industry.}		
In terms of experience, we split the participants into three groups: \begin{itemize}
    \item Less experienced - 19\% have between 0 and 2 years of experience;
    \item Experienced - 47\% have between 3 and 7 years; 
    \item Very experienced - 34\% reported more than 8 years of experience. 
\end{itemize}

    \item \textit{Roles.}
    20\% are Jr. SQA Engineers (level I-II), 23\%	are SQA Engineers (level III), 41\% are Senior SQA Engineers (level IV) or SQA specialists, 11\% are SQA Team Leaders or SQA Managers, and 5\% are SQA Engineering Coordinator (Head of SQA). Around 26\% usually perform more than one role.
    
    \item \textit{SQA Activities.}
    The activities that the SQA engineers perform every day or many times a week include: creating and executing manual tests, creating and maintaining automated test scripts, defining metrics to measure the quality of the product, reporting bugs in the code, sharing knowledge with the team, reviewing test cases (manual or automated), creating test plans, studying tools to be considered for adoption by the project, and helping to define user story acceptance criteria.

    \item \textit{Types of Testing.}
    The types of testing that the SQA engineers mostly perform at work include regression (66\%), end-to-end (64\%), integration (62\%), interface (51\%), and usability testing (43\%); the testing performed by smaller groups of SQAs were: load, stress, performance, security, sanity, and unit testing.
\end{itemize}

\section{Results}
\label{results}

\begin{figure*}[!htp]
\centering{\includegraphics[scale=0.6]{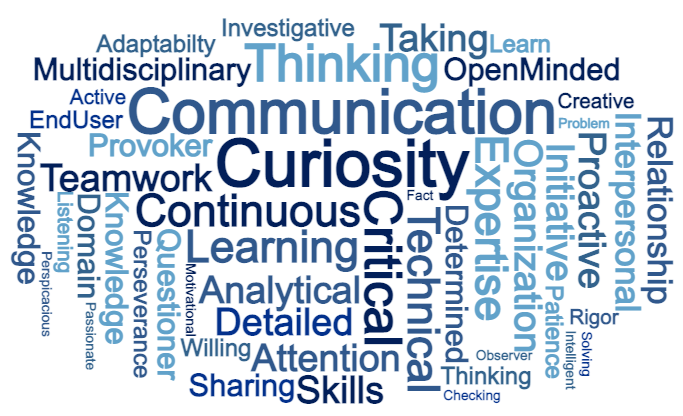}}
\caption{Word cloud of the attributes mentioned by the participants where the frequency of each word is shown with font \\size. It was created using a free online word generator \cite{clouds}.
}
\label{wordcloud}
\end{figure*} 

In this section, we first present the word cloud consisting of all the words used by study participants. 
We then discuss each of the three research questions mentioned in Section \ref{intro}.

\subsection{RQ1 - What are the attributes of great SQA Engineers? Why are these attributes viewed as important?}

A quick and informal view of importance is shown in Figure \ref{wordcloud}. It gives us a quick overview of the most frequent words used by the participants in the questionnaire and during the interviews. For instance, \textit{curiosity} was mentioned by 20 out of 25 respondents, while \textit{patience} was cited by 4.

Performing a coding procedure allowed us to uncover 25 SQA engineers' attributes and group them into five categories: management, decision-making, technical, social, and personal. 

\subsubsection{\textbf{Management}}
The management category groups three attributes that help manage the SQA daily tasks: time, priority,  and people.

\begin{itemize}
    \item \textbf{Time management}.
    Time management refers to the ability to use the working hours efficiently. Since the demand can change several times in a day, a good way to manage the time well is to prioritize the features to be tested and estimate the testing effort (\#P12). A great SQA should know where to look for information to understand the business rules (\#P1). \textit{``Time management is definitively important for an SQA engineer, in particular, because quite often there is not a lot of time [to complete the tests], and a lot to get done"} (\#P07).

    \item \textbf{Priority Management}. 
    It refers to the ability to work with several tasks at the same time, knowing how to assess the risks and impacts of each change and prioritizing each task as needed. It is common for an SQA engineer to split their attention on more than one task at the same time. The size of a team that follows an agile methodology usually varies per company, but we noticed that the number of developers is always higher than the number of SQAs. For instance, \#P20 stated that he was the only SQA engineer in the team at his previous job. For other respondents like \#P17 and \#P23, it was a 4:1 developers to SQA ratio. Therefore, a great SQA engineer should know how to assess the priorities and risks of each change and manage when to bounce from one task to another. Examples of frequent emergency tasks are production issues that the SQA needs to help troubleshoot and replicate on the lower environments; an unexpected failure, or simply a new project with a higher business value. \textit{``This ends up being a little stressful because it always comes like this: "I need it yesterday"} (\#P05). 

    \item \textbf{Leadership}. 
    Five participants (20\%) mentioned that it is important for an SQA engineer to have leadership skills. This is about acting as a leader even without the title by motivating coworkers to do their best work and considering the ethical implications of their decisions and actions. An SQA with this characteristic would guide the team and care about the performance of the entire team even when she/he is not responsible for the team. \#P3 mentioned that \textit{``there are people in his team that ``wear" the manager hat when needed and pass confidence to the entire team"}. An SQA team lead highlighted the following: \textit{``There is someone on my team that works as a secondary lead when I am not available. He is a backup for me, and this makes him a great SQA"} (\#P19).
\end{itemize}

\subsubsection{\textbf{Decision Making}}

This category considers the relative impacts and benefits of potential actions to choose the most appropriate ones. Participants mentioned some examples of the SQA’s ability to base a decision on multiple aspects. We split them into four attributes that we discuss in more detail next.

\begin{itemize}
\item \textbf{Analytic and systematic view}. It refers to the ability to perform root-cause and risk analyses. For example, a great SQA engineer should know whether to approve a release or push it back until an issue is resolved because she/he knows the impact of the change (\#P1). 

\item \textbf{Critical Thinking}. 
It refers to the analysis of the facts to form a perception of something. An SQA engineer needs to perform a root cause analysis to identify the main causes of a problem and propose solutions (\#P2).
An SQA that thinks critically can propose brainstorming discussions to confront ideas, get insights, and challenge common assumptions. Also, she/he can make important decisions every day based on her/his critical thinking during the testing process (\#P14), and it will impact positively the effectiveness of the tests (\#P12).

\item \textbf{Holistic view}. It is the ability to see the system as a whole and address problems in an organized, and conscientious manner. 
The known term ``thinking outside the box" is very common in SQA, in the sense of asking questions such as \textit{``what am I testing?", ``why am I doing this?", ``what else could be affected by this change?"} (\#P3), \textit{``what are the edge cases?", ``what should be the expected results?"} (\#P5). To answer those types of questions, the SQA engineer needs to have a broad view of the business rules (\#P6), needs to consider testing the integration with related systems, and not be limited by its supposed constraints (\#P7).

\item \textbf{Being customer-centric} is the ability to always be aligned with the customer so that their expectations are more likely to be met.
It is about thinking not just like a common SQA: ``\textit{I am going to test this product here and that's it}", but ``\textit{What value am I delivering? Why am I testing this?}". It is about adding value to work (\#P5). By understanding what you are doing, you will do a much better job (\#P03).
According to \#P01, the SQA engineer has to focus on whether this functionality will be useful for the end users, whether will they understand how it works, and how to use it. 
\end{itemize} 

\subsubsection{\textbf{Technical}}
This category represents technical skills that are useful in an SQA position.

\begin{itemize}

\item \textbf{Domain Knowledge}.
It refers to understanding the environment in which the IT system operates, encompassing the understanding of the company's product values and knowing the rules and constraints to help build the test plans. Since the SQA professional needs to talk with different stakeholders, from the developer to the customer, she/he will most likely understand the context of the functionality (\#P3) and know the business rules involved (\#P4). It impacts the testing process directly because it allows the SQA to understand whether a change makes sense or not and analyze the impacts caused by a single change (\#P9).

\item \textbf{Technical expertise}.
Technical expertise refers to holding several abilities that can facilitate daily tasks. For instance, being familiar with and comfortable using test case management tools, code deployment management, and bug trackers. Also, the ability to look for a log in the terminal, and have experience with Structured Query Language (SQL) to generate, manage, and use data for a test case. Another example is understanding related areas such as DevOps, database, and product (\#P5), and not just focusing on software quality knowledge (\#P8). Finally, \#P10 stated that \textit{``an SQA engineer with technical expertise knows a little about each thing, making it easier for her/him to become an expert in anything"}.

\item \textbf{Programming skills}.
It is the ability to help read, understand, and debug a development code, review pull requests with confidence, and ability to code if needed. Forty percent of the participants mentioned that development skills positively impact their testing. Some advantages are that they will be able to understand the code and take a look at the pull request to assess the risks (\#P13); even when they do not look at the code, they can think of several testing scenarios based on a more technical conversation with developers and architects (\#P16). In addition, participating in the entire development process is necessary to acquire more technical skills (\#P13). 
\end{itemize}

Another attribute grouped into this category was having \textit{theoretical knowledge of the SQA area} to know the best test strategies and apply them to each test scenario.

\subsubsection{\textbf{Social}}
This category groups the attributes related to how SQA engineers relate with the team and other SQAs.

\begin{itemize}

\item \textbf{Communication.}
This is about having the ability to communicate effectively with all stakeholders when asking questions or proposing something to others (\#P2). The SQA engineer often acts as a bridge between the customer/product owner and the developer, so it is important to understand what the customer is asking for (\#P13) and know all the impacts of a change (\#P6). The SQA engineer needs to adopt different languages to speak to people in different roles, that do not understand in the same way (\#P14). When reporting a bug, for instance, the SQA needs to be careful because she/he is dealing with other people's mistakes (\#P10). 

\item \textbf{Share knowledge with others}.
It refers to the exchange of knowledge, skills, and experiences with others (\#P11). This is about sharing knowledge with others (\#P9), documenting procedures to make the information available to everyone, or promoting training when needed. 

\item \textbf{Collaborate with others}.
Another attribute within this category was collaboration, which refers to supporting and guiding the team and being active in the SQA community. The respondent \#P9 mentioned that a good way to pass on knowledge is to promote content at work and outside of work (e.g., on LinkedIn).

\end{itemize}

\subsubsection{\textbf{Personal}}
This category groups the most important soft skills of a great SQA engineer. 

\begin{itemize}
\item \textbf{Adaptability}. It refers to the ability to change her/his approach to doing things to suit a new situation. The respondent \#P1 mentioned that the information technology (IT) area is constantly changing due to new technologies and tools being released every day, and SQA engineers need to adapt themselves to the market (\#P5). At the same time, they need to be flexible and know how to deal with changes in priorities (\#P11).

\item \textbf{Continuous learning}.
According to \#P5, there is always room for improvement, and being a continuous learner is about it. The respondent \#P6 mentioned that great SQA engineers always challenge themselves by building new skills and stepping out of their comfort zones. The respondent \#P14 suggested that setting up personal and professional milestones is a good way to keep up to date with learning new technologies. 

\item \textbf{Curiosity}. It is about employee engagement and agility, in the sense of exploring and going beyond the basics (\#P6) and not being satisfied with the traditional ways of testing (\#P11). A curious SQA engineer will want to understand everything, question the requirements, and so on (\#P9). When a bug is found, the curious SQA will likely analyze the root causes to understand what went wrong (\#P2) and implement improvements (\#P5). Finally, it is about looking for things that are not functioning correctly and searching for ways to improve them (\#P16).

\item \textbf{Resilient}. It includes characteristics such as resilience (\#P1, \#P6), attitude (\#P2), and serenity (\#P5), among others. 
Some participants shared that there is still resistance from some developers to acknowledge an error in their code. Determined SQA engineers have their vision of the problem (\#P14) and persist with their intuitions and decisions (\#P13).

\item \textbf{Courageous}. It is courageous to say ``no" to allowing things to be approved for deployment in production if it is not fully tested (\#P7).

\item \textbf{Detail orientated}. 
It is about paying attention to minimum details when planning, writing, and executing a test case (\#P9). The respondent \#P10 stated that when someone writes a test case or documentation, this person does not own it. Everyone who has access to the document should be able to understand it well and reproduce all test steps without any further questions. Besides, a great SQA engineer can detect a problem before starting testing, just by thinking about the implicit details involved with the functionality, such as potential feature interactions that may occur (\#P14). 

\item \textbf{Proactive}.
There are several opportunities for the SQA professional to be proactive. When taking ownership of a project voluntarily (\#P9), having a critical sense, and seeking more information when needed (\#P13), do not be accommodated or stop at the dependencies that may arise, like waiting on a developer to fix an issue (\#P14), and always looking to improve the testing process (\#P15). 

\end{itemize}

Other attributes within this category are \textit{being creative, perfectionist, suspicious, and patient}.

\begin{myframe}{Observation 1}
        Curiosity and continuous learning \textit{(personal)}; communication and teamwork (collaborate with others) \textit{(social)}; critical thinking and analytical view (decision-making), technical expertise and domain knowledge \textit{(technical)}; time and priority \textit{(management)} were the most frequent attributes identified through the interviews.
\end{myframe}

\subsection{RQ2 - How does the SQA community perceive the importance of these attributes?}
\label{results2}

\begin{figure*}[!btp]
\centering{\includegraphics[width=\linewidth]
{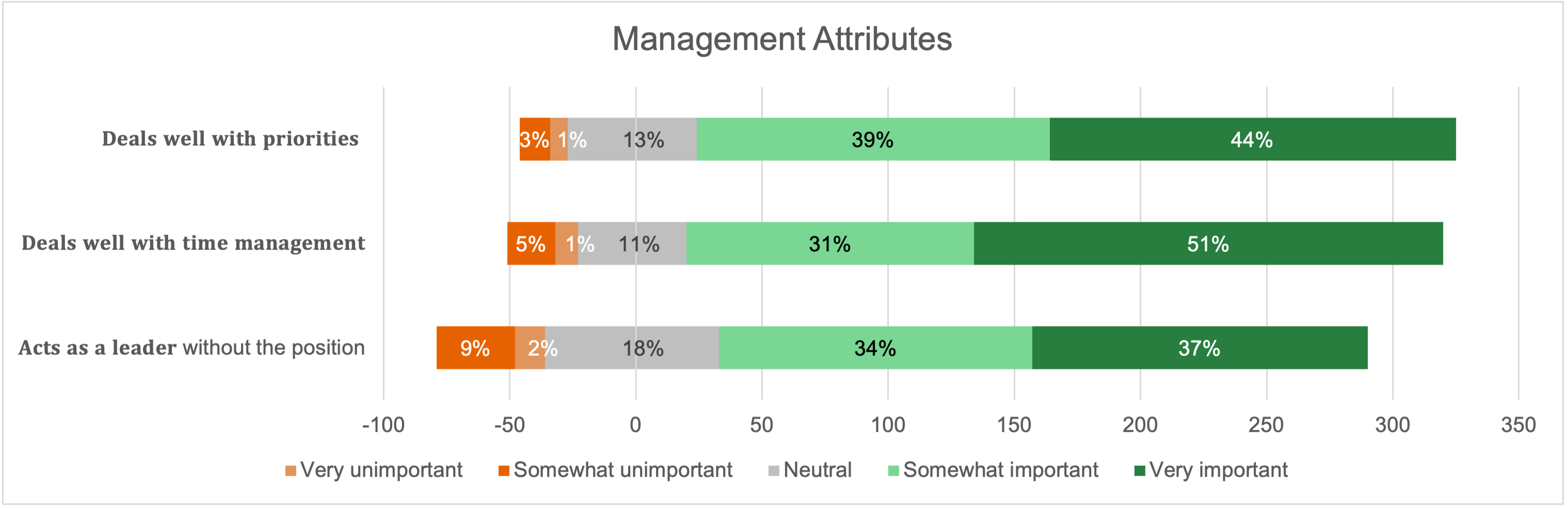}}
\caption{Importance of Management Attributes reported by the respondents}
\label{management1}
\end{figure*}

\begin{table*}[!tb]
\centering
\caption{Results of the rating survey for the management category per gender}
\label{tab:managementGender}
\begin{tabular}{@{}lccccc@{}}
\cmidrule(l){2-6}
                    & \multicolumn{5}{c}{\textbf{Female}}                                                     \\ \cmidrule(l){2-6} 
                    & Very important & Somewhat important & Neutral & Somewhat unimportant & Very unimportant \\
Time management     & 63.2\%         & 31.6\%             & 4.5\%   & 0.8\%                & 0.0\%            \\
Priority management & 55.6\%         & 32.3\%             & 10.5\%  & 1.5\%                & 0.0\%            \\
Act as a leader     & 41.4\%         & 36.8\%             & 15.0\%  & 6.8\%                & 0.0\%            \\ \cmidrule(l){2-6} 
                    & \multicolumn{5}{c}{\textbf{Male}}                                                       \\ \cmidrule(l){2-6} 
                    & Very important & Somewhat important & Neutral & Somewhat unimportant & Very unimportant \\
Time management     & 44.3\%         & 31.1\%             & 14.9\%  & 7.9\%                & 1.8\%            \\
Priority management & 38.2\%         & 41.7\%             & 14.9\%  & 4.4\%                & 0.9\%            \\
Acts as a leader    & 33.3\%         & 32.9\%             & 20.6\%  & 9.2\%                & 3.9\%           
\end{tabular}%
\end{table*}

\begin{table*}[!bt]
\caption{Results of the rating survey for the management category per experience level}
\label{tab:managementExp}
\centering
\begin{tabular}{lccccc}
\cline{2-6}
                    & \multicolumn{5}{c}{\textbf{Less Experienced}}                                           \\ \cline{2-6} 
                    & Very important & Somewhat important & Neutral & Somewhat unimportant & Very unimportant \\
Priority management & 54.4\%         & 33.8\%             & 7.4\%   & 2.9\%                & 1.5\%            \\
Time management     & 51.5\%         & 38.2\%             & 7.4\%   & 1.5\%                & 1.5\%            \\
Acts as a leader    & 38.2\%         & 33.8\%             & 22.1\%  & 4.4\%                & 1.5\%            \\ \cline{2-6} 
                    & \multicolumn{5}{c}{\textbf{Experienced}}                                                \\ \cline{2-6} 
                    & Very important & Somewhat important & Neutral & Somewhat unimportant & Very unimportant \\
Time management     & 54.1\%         & 30.2\%             & 9.9\%   & 5.2\%                & 0.6\%            \\
Priority management            & 44.8\%         & 37.8\%             & 14.0\%  & 3.5\%                & 0.0\%            \\
Acts as a leader    & 36.0\%         & 32.0\%             & 19.8\%  & 9.3\%                & 2.9\%            \\ \cline{2-6} 
                    & \multicolumn{5}{c}{\textbf{Very Experienced}}                                           \\ \cline{2-6} 
                    & Very important & Somewhat important & Neutral & Somewhat unimportant & Very unimportant \\
Time management     & 47.2\%         & 29.3\%             & 14.6\%  & 6.8\%                & 1.5\%            \\
Priority management & 38.2\%         & 42.3\%             & 15.4\%  & 3.0\%                & 0.8\%            \\
Acts as a leader    & 35.8\%         & 37.4\%             & 14.6\%  & 9.0\%                & 2.3\%           
\end{tabular}%
\end{table*}

The twenty-five attributes that emerged from the interviews were brought into the survey and divided into the following categories: management, decision-making, technical, social, and personal. This section presents the level of importance of each attribute based on the survey's results. For each category, we analyze the results of the whole group and the results of the rating survey per gender and experience level. 

\subsubsection{\textbf{Management Attributes}}

As we can see in Figure \ref{management1}, \textit{dealing with priorities} was considered as most important by the respondents. A great SQA engineer knows how to assess the risks and impacts of each change even when working with several tasks at the same time. Only 3\% considered it as very unimportant, while 39\% considered it as somewhat important and 44\% as very important; followed by \textit{time management} where 51\% considered it as very important.

\textbf{Contextual factor - Gender}\\
Looking at gender, although the order of importance seemed to not change between females and males, we could notice a few differences: 63.2\% of the women classified \textit{time management} as ``Very Important" while men were 44.3\%. As we can see in Table \ref{tab:managementGender}, men tended to be more neutral in their opinion than women; and unlike the male gender, female participants did not classify any management attribute as ``Very Unimportant".

\textbf{Contextual factor - Experience}\\
Considering the experience level as an SQA engineer in the management category (see Table \ref{tab:managementExp}), we can notice
that \textit{acting as a leader without the title} seems to not be as important as \textit{dealing with time management} and \textit{dealing with priority changes}. Also, comparing the levels of experience we notice that \textit{dealing with priorities} is more frequent but as the SQA becomes more experienced, \textit{time management} becomes more critical. A possible reason is that they have to do mentoring, do their tasks, talk to people, and so on.

\begin{myframe}{Observation 2}
\textit{Dealing with priorities} and \textit{time management} had a similar level of importance among the respondents, but as the SQA becomes more experienced, \textit{time management} becomes more critical and gains more importance.
\end{myframe}

\subsubsection{\textbf{Decision-making Attributes}}

\begin{figure*}[!htp]
\centering{\includegraphics[width=\linewidth]{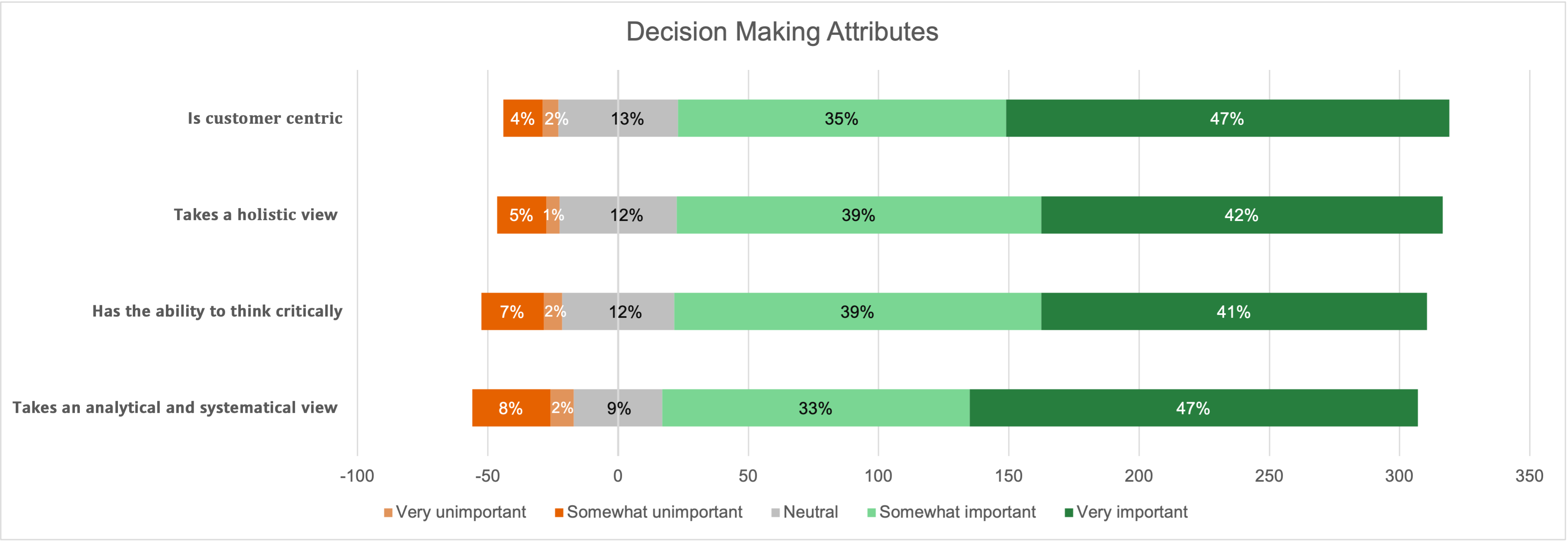}}
\caption{Importance of Decision-Making attributes reported by the respondents}
\label{decision1}
\end{figure*}

\begin{table*}[!hbt]
\caption{Results of the rating survey for the decision-making category per gender}
\label{tab:decisionGender}
\begin{tabular}{@{}lccccc@{}}
\cmidrule(l){2-6}& \multicolumn{5}{c}{\textbf{Female}}  \\ \cmidrule(l){2-6}  & Very important & Somewhat important & Neutral & Somewhat unimportant & Very unimportant \\
Is customer-centric                 & 55.6\%         & 33.1\%             & 9.8\%   & 1.5\%                & 0.0\%            \\
Analytical and systematical view    & 54.1\%         & 27.8\%             & 10.5\%  & 6.0\%                & 1.5\%            \\
Takes a holistic view               & 52.6\%         & 33.1\%             & 11.3\%  & 3.0\%                & 0.0\%            \\
Has the ability to think critically & 49.6\%         & 36.1\%             & 8.3\%   & 5.3\%                & 0.8\%            \\ \cmidrule(l){2-6} 
                                    & \multicolumn{5}{c}{\textbf{Male}}                                                       \\ \cmidrule(l){2-6} 
                                    & Very important & Somewhat important & Neutral & Somewhat unimportant & Very unimportant \\
Analytical and systematical view    & 43.4\%         & 35.1\%             & 8.8\%   & 9.6\%                & 3.1\%            \\
Is customer-centric                 & 41.7\%         & 35.5\%             & 14.5\%  & 5.7\%                & 2.6\%            \\
Takes a holistic view               & 36.0\%         & 42.1\%             & 13.2\%  & 6.6\%                & 2.2\%            \\
Has the ability to think critically & 35.5\%         & 40.4\%             & 14.0\%  & 7.5\%                & 2.6\%           
\end{tabular}%
\end{table*}

\begin{table*}[!hbt]
\caption{Results of the rating survey for the decision-making category per experience level}
\label{tab:decisionExp}
\begin{tabular}{@{}lccccc@{}}
\cmidrule(l){2-6}
                                 & \multicolumn{5}{c}{\textbf{Less Experienced}}                                           \\ \cmidrule(l){2-6} 
                                 & Very important & Somewhat important & Neutral & Somewhat unimportant & Very unimportant \\
Analytical and systematical view & 52.9\%         & 35.3\%             & 5.9\%   & 2.9\%                & 2.9\%            \\
Think critically                 & 52.9\%         & 33.8\%             & 5.9\%   & 2.9\%                & 4.4\%            \\
Is customer-centric              & 45.6\%         & 36.8\%             & 13.2\%  & 2.9\%                & 1.5\%            \\
Takes a holistic view            & 39.7\%         & 47.1\%             & 7.4\%   & 2.9\%                & 2.9\%            \\ \cmidrule(l){2-6} 
                                 & \multicolumn{5}{c}{\textbf{Experienced}}                                                \\ \cmidrule(l){2-6} 
                                 & Very important & Somewhat important & Neutral & Somewhat unimportant & Very unimportant \\
Analytical and systematical view & 46.5\%         & 29.1\%             & 9.9\%   & 11.0\%               & 3.5\%            \\
Holistic view                    & 45.3\%         & 33.1\%             & 15.7\%  & 5.8\%                & 0.0\%            \\
Is customer-centric              & 44.8\%         & 35.5\%             & 12.2\%  & 5.8\%                & 1.7\%            \\
Think critically                 & 34.9\%         & 45.9\%             & 11.0\%  & 7.0\%                & 1.2\%            \\ \cmidrule(l){2-6} 
                                 & \multicolumn{5}{c}{\textbf{Very Experienced}}                                           \\ \cmidrule(l){2-6} 
                                 & Very important & Somewhat important & Neutral & Somewhat unimportant & Very unimportant \\
Is customer-centric              & 50.4\%         & 32.5\%             & 13.0\%  & 2.4\%                & 1.6\%            \\
Analytical and systematical view & 45.5\%         & 35.8\%             & 10.6\%  & 7.3\%                & 0.8\%            \\
Think critically                 & 42.3\%         & 31.7\%             & 16.3\%  & 8.1\%                & 1.6\%            \\
Takes a holistic view            & 39.8\%         & 41.5\%             & 10.6\%  & 5.7\%                & 2.4\%           
\end{tabular}%
\end{table*}

If we only consider the ``Most Important" scale for the \textit{decision making} category, we notice that being a \textit{customer centric} SQA, focusing on the intersection between the needs of the user, and \textit{taking an analytical and systematical view} was chosen by almost half of the participants. However, as we can see in Figure \ref{decision1}, considering all the scales, the most important decision-making attributes were: a \textit{customer centric} SQA engineer who \textit{takes a holistic view.}

\textbf{Contextual factor - Gender} \\ 
It is interesting to look into gender for this category because we can see a slight difference between the two most classified attributes for this category. Table \ref{tab:decisionGender} shows that while the \textit{female} participants think that being a customer-centric SQA is more important, \textit{taking an analytical and systematical view} is considered more important for the male participants although we could not find a possible reason for this during the interviews and survey. 

\textbf{Contextual factor - Experience}\\
Splitting the group of participants between levels of experience, we notice that \textit{less experienced}  SQA engineers think that a great SQA needs to \textit{take an analytical and systematical view}, and as they become more \textit{experienced}, the SQAs think that being \textit{customer centric} is the most important attribute for this category (see Table \ref{tab:decisionExp}).

\begin{myframe}{Observation 3}
The decision-making attributes considered very important by the respondents were: being \textit{customer centric} and \textit{taking an analytical and systematical view}. As they become more experienced, being a \textit{customer-centric} SQA becomes more important.
\end{myframe}

\subsubsection{\textbf{Technical Attributes}}

 \begin{figure*}[!htp]
\centering{\includegraphics[width=\linewidth]{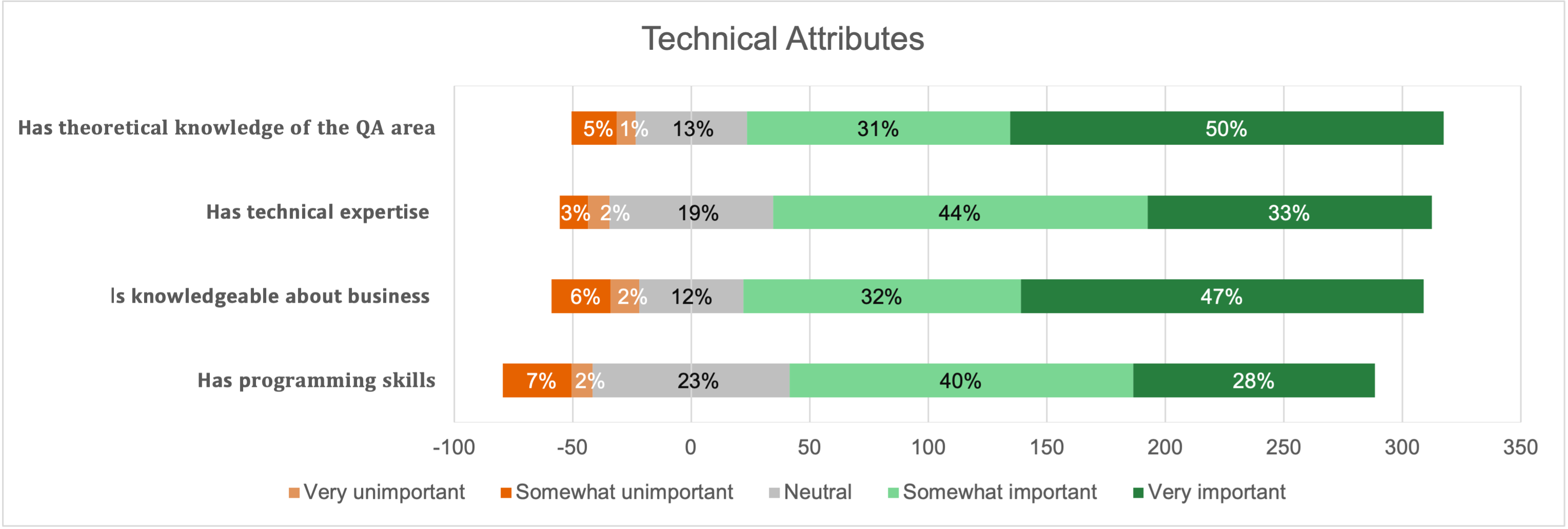}}
\caption{Importance of Technical Attributes reported by the respondents}
\label{technical1}
\end{figure*}

\begin{table*}[!tb]
\centering
\caption{Results of the rating survey for the technical category per gender}
\label{tab:technicalGender}
\begin{tabular}{lccccc}
\cline{2-6} & \multicolumn{5}{c}{\textbf{Female}}     \\ \cline{2-6} & Very important & Somewhat important & Neutral & Somewhat unimportant & Very unimportant \\
Has theoretical knowledge of the SQA area & 61.7\%         & 27.1\%             & 8.3\%   & 2.3\%                & 0.8\%            \\
Is Knowledgeable about domain          & 57.9\%         & 29.3\%             & 8.3\%   & 4.5\%                & 0.0\%            \\
Has technical expertise                  & 40.6\%         & 42.1\%             & 15.8\%  & 1.5\%                & 0.0\%            \\
Has programming skills                   & 30.8\%         & 44.4\%             & 18.8\%  & 6.0\%                & 0.0\%            \\ \cline{2-6} & \multicolumn{5}{c}{\textbf{Male}}   \\ \cline{2-6}  & Very important & Somewhat important & Neutral & Somewhat unimportant & Very unimportant \\
Has theoretical knowledge of the SQA area & 43.4\%         & 32.9\%             & 15.8\%  & 6.1\%                & 1.8\%            \\
Is Knowledgeable about domain          & 40.8\%         & 33.3\%             & 14.5\%  & 7.5\%                & 3.9\%            \\
Has technical expertise                  & 28.5\%         & 44.3\%             & 21.1\%  & 3.5\%                & 2.6\%            \\
Has programming skills                   & 25.9\%         & 37.7\%             & 25.4\%  & 8.3\%                & 2.6\%           
\end{tabular}
\end{table*}

\begin{table*}[!tp]
\caption{Results of the rating survey for the technical category per experience level}
\label{tab:techicalExp}
\begin{tabular}{@{}lccccc@{}}
\cmidrule(l){2-6}
                                         & \multicolumn{5}{c}{\textbf{Less Experienced}}                                           \\ \cmidrule(l){2-6} 
                                         & Very important & Somewhat important & Neutral & Somewhat unimportant & Very unimportant \\
Has theoretical knowledge of the SQA area & 60.3\%         & 32.4\%             & 4.4\%   & 0.0\%                & 2.9\%            \\
Is Knowledgeable about domain          & 54.4\%         & 36.8\%             & 4.4\%   & 2.9\%                & 1.5\%            \\
Has technical expertise                  & 38.2\%         & 48.5\%             & 11.8\%  & 0.0\%                & 1.5\%            \\
Has programming skills                   & 35.3\%         & 38.2\%             & 22.1\%  & 2.9\%                & 1.5\%            \\ \cmidrule(l){2-6} 
                                         & \multicolumn{5}{c}{\textbf{Experienced}}                                                \\ \cmidrule(l){2-6} 
                                         & Very important & Somewhat important & Neutral & Somewhat unimportant & Very unimportant \\
Has theoretical knowledge of the SQA area & 51.7\%         & 31.4\%             & 11.6\%  & 4.7\%                & 0.6\%            \\
Is Knowledgeable about domain          & 47.7\%         & 27.9\%             & 11.6\%  & 9.9\%                & 2.9\%            \\
Has technical expertise                  & 32.0\%         & 44.2\%             & 19.2\%  & 2.3\%                & 2.3\%            \\
Has programming skills                   & 30.8\%         & 34.9\%             & 24.4\%  & 9.3\%                & 0.6\%            \\ \cmidrule(l){2-6} 
                                         & \multicolumn{5}{c}{\textbf{Very Experienced}}                                           \\ \cmidrule(l){2-6} 
                                         & Very important & Somewhat important & Neutral & Somewhat unimportant & Very unimportant \\
Has theoretical knowledge of the SQA area & 43.1\%         & 28.5\%             & 19.5\%  & 7.3\%                & 1.6\%            \\
Is Knowledgeable about domain          & 41.5\%         & 35.8\%             & 17.1\%  & 3.3\%                & 2.4\%            \\
Has technical expertise                  & 31.7\%         & 39.8\%             & 22.8\%  & 4.9\%                & 0.8\%            \\
Has programming skills                   & 20.3\%         & 48.0\%             & 21.1\%  & 7.3\%                & 3.3\%           
\end{tabular}%
\end{table*}
Having technical competency in the SQA profession is not usually required, but it was much discussed among the participants during the interviews as a good thing to have. In the rating survey, we were able to see which technical attributes are viewed as more important for the SQA community.
According to Figure \ref{technical1}, having \textit{theoretical knowledge of the SQA area}, which refers to knowing the best test strategies to apply to each test scenario, was considered the most important technical attribute. Followed by having \textit{technical expertise}, which refers to having the ability to look for a log, SQL skills, Linux basic commands skills, and so on. The next attribute in the priority list was being \textit{knowledgeable about the domain}, which refers to understanding the product’s value and knowing the rules and constraints to help build the test plans; and the last on the importance scale was having \textit{programming skills}, which is the ability to help read, understand and debug a development code, review pull requests with confidence, and ability to code if needed.

\textbf{Contextual factor - Gender}\\
Even after splitting the group by gender, there is no noticeable difference in the order of priority as can be seen in Table \ref{tab:technicalGender}. For both genders, having \textit{theoretical knowledge of the SQA area} is considered the most important attribute, and having \textit{programming skills} is considered the less important technical attribute for both groups. 

 \begin{figure*}[!bht]
\centering{\includegraphics[width=\linewidth]{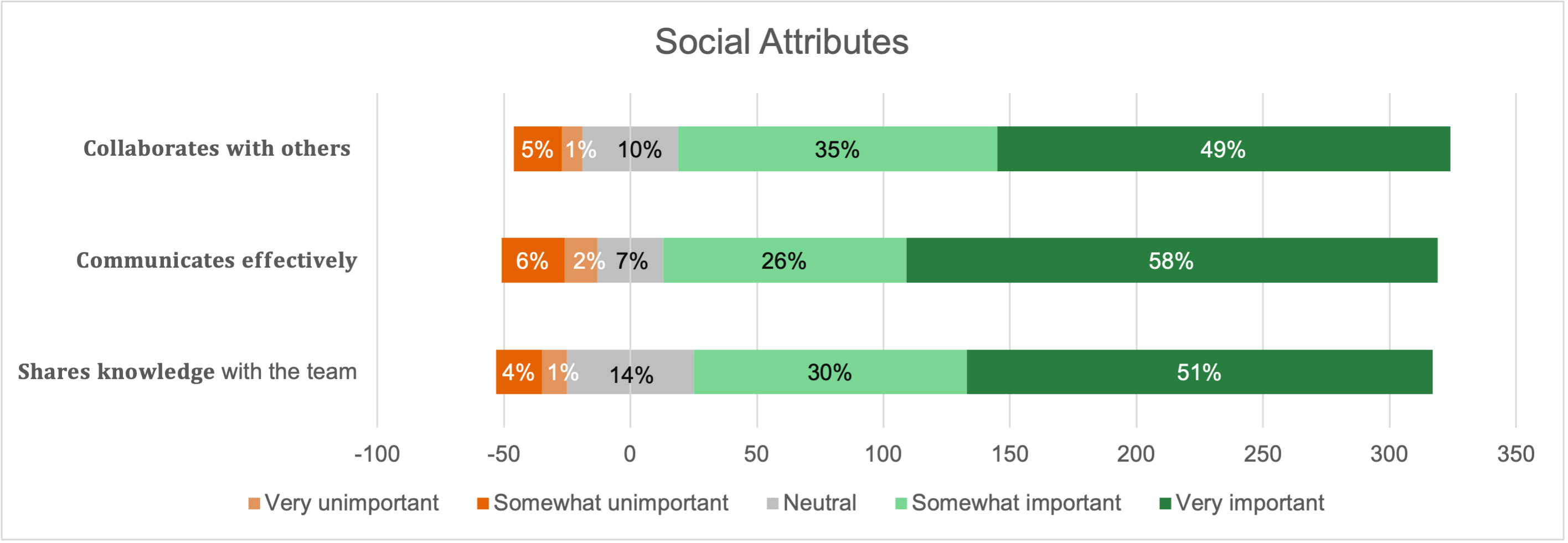}}
\caption{Importance of Social Attributes reported by the respondents}
\label{social1}
\end{figure*}

\textbf{Contextual factor - Experience}\\
Similarly, looking at Table \ref{tab:techicalExp}, we notice that different levels of experience as SQA engineers seem to not make a difference in the order of priority for technical attributes. Having \textit{theoretical knowledge of the SQA area} is considered the most important attribute and having \textit{programming skills} is considered the less important technical attribute for both groups.

\begin{table*}[!htp]
\caption{Results of the rating survey for the social category per gender}
\label{tab:socialGender}
\centering
\begin{tabular}{@{}lccccc@{}}
\cmidrule(l){2-6}
 &
  \multicolumn{5}{c}{\textbf{Female}} \\ \cmidrule(l){2-6} 
 &
  \multicolumn{1}{l}{Very important} &
  \multicolumn{1}{l}{Somewhat important} &
  \multicolumn{1}{l}{Neutral} &
  \multicolumn{1}{l}{Somewhat unimportant} &
  \multicolumn{1}{l}{Very unimportant} \\
Communicates effectively &
  69.9\% &
  21.8\% &
  5.3\% &
  3.0\% &
  0.0\% \\
Collaborates with others &
  61.7\% &
  29.3\% &
  7.5\% &
  1.5\% &
  0.0\% \\
Shares knowledge with others &
  61.7\% &
  23.3\% &
  12.8\% &
  2.3\% &
  0.0\% \\ \cmidrule(l){2-6} 
 &
  \multicolumn{5}{c}{\textbf{Male}} \\ \cmidrule(l){2-6} 
 &
  \multicolumn{1}{l}{Very important} &
  \multicolumn{1}{l}{Somewhat important} &
  \multicolumn{1}{l}{Neutral} &
  \multicolumn{1}{l}{Somewhat unimportant} &
  \multicolumn{1}{l}{Very unimportant} \\
Shares knowledge with others &
  50.4\% &
  29.4\% &
  8.3\% &
  8.3\% &
  3.5\% \\
Communicates effectively &
  44.7\% &
  32.9\% &
  14.5\% &
  5.7\% &
  2.2\% \\
Collaborates with others &
  42.5\% &
  37.3\% &
  12.3\% &
  6.6\% &
  1.3\% \\ \bottomrule
\end{tabular}%
\end{table*}

\begin{table*}[!htb]
\caption{Results of the rating survey for the social category per experience level}
\label{tab:socialExp}
\centering
\begin{tabular}{@{}lccccc@{}}
\cmidrule(l){2-6}
 &
  \multicolumn{5}{c}{\textbf{Less Experienced}} \\ \cmidrule(l){2-6} 
 &
  Very important &
  Somewhat important &
  Neutral &
  Somewhat unimportant &
  Very unimportant \\
Communicates effectively &
  67.6\% &
  22.1\% &
  4.1\% &
  1.5\% &
  1.5\% \\
Shares  knowledge with others &
  60.3\% &
  27.9\% &
  4.9\% &
  1.5\% &
  1.5\% \\
Collaborates with others &
  51.5\% &
  38.2\% &
  2.4\% &
  2.9\% &
  2.9\% \\ \cmidrule(l){2-6} 
 &
  \multicolumn{5}{c}{\textbf{Experienced}} \\ \cmidrule(l){2-6} 
 &
  Very important &
  Somewhat important &
  Neutral &
  Somewhat unimportant &
  Very unimportant \\
Communicates effectively &
  57.0\% &
  25.0\% &
  5.2\% &
  9.9\% &
  2.9\% \\
Collaborates with others &
  51.2\% &
  32.6\% &
  11.6\% &
  4.7\% &
  0.0\% \\
Shares  knowledge with others &
  50.0\% &
  29.7\% &
  14.5\% &
  5.2\% &
  0.6\% \\ \cmidrule(l){2-6} 
 &
  \multicolumn{5}{c}{\textbf{Very Experienced}} \\ \cmidrule(l){2-6} 
 &
  \multicolumn{1}{l}{Very important} &
  \multicolumn{1}{l}{Somewhat important} &
  \multicolumn{1}{l}{Neutral} &
  \multicolumn{1}{l}{Somewhat unimportant} &
  \multicolumn{1}{l}{Very unimportant} \\
Communicates effectively &
  \multicolumn{1}{l}{53.7\%} &
  \multicolumn{1}{l}{30.9\%} &
  \multicolumn{1}{l}{9.8\%} &
  \multicolumn{1}{l}{4.1\%} &
  \multicolumn{1}{l}{1.6\%} \\
Shares  knowledge with others &
  \multicolumn{1}{l}{46.3\%} &
  \multicolumn{1}{l}{30.9\%} &
  \multicolumn{1}{l}{15.4\%} &
  \multicolumn{1}{l}{4.9\%} &
  \multicolumn{1}{l}{2.4\%} \\
Collaborates with others &
  \multicolumn{1}{l}{45.5\%} &
  \multicolumn{1}{l}{35.8\%} &
  \multicolumn{1}{l}{12.2\%} &
  \multicolumn{1}{l}{5.7\%} &
  \multicolumn{1}{l}{0.8\%} \\ \bottomrule
\end{tabular}%
\end{table*}

\begin{myframe}{Observation 4} 
Having \textit{theoretical knowledge of the SQA area} was considered the most important attribute of the technical category regardless of the contextual factors of gender and experience.
\end{myframe}

\subsubsection{\textbf{Social Attributes}}

As shown in Figure \ref{social1}, all three social attributes were viewed as strongly positive among the participants. The attribute \textit{collaborates with others} by supporting the team, being active in the SQA community can be highlighted as most important by the respondents with 49\% as very important, 35\% as somewhat important, only 10\% were neutral, and only 5\% considered as somewhat unimportant and 1\% as very unimportant. The attribute \textit{communicates effectively} when reporting a bug, when asking questions, or when dealing with different stakeholders had the highest percentage as very important (58\%) and 26\% as somewhat important. Finally, the attribute \textit{share knowledge with others} was rated with 51\% as very important and 30\% as somewhat important.

\textbf{Contextual factor - Gender}\\
As we can notice in Table \ref{tab:socialGender}, the majority of the female participants (almost 70\%) considered the attribute \textit{communicate effectively} as very important. Followed by the attributes: \textit{collaborate with others} (61.7\%) and then \textit{share knowledge with others} (61.7\%). The male group seemed to have a more divided opinion about this category, where 50\% of the men think that \textit{share knowledge with others} is very important and 44.7\% think that \textit{communicates effectively} is very important. The attribute \textit{collaborates with others} only had 42.5\% votes as ``very important".
When looking at gender for the social category, we can notice that women think that communicating effectively is key for a great SQA engineer while men think that sharing knowledge with the team is more critical. 

\textbf{Contextual factor - Experience}\\
Splitting the survey data of the social category among levels of experience does not seem to impact much. The attribute \textit{communicating effectively} is highlighted as most important by all three levels of experience. Table \ref{tab:socialExp} highlights that for the less experienced group, the attribute \textit{collaborates with others} is in third place. One potential reason for this is that it might be hard for an SQA junior (having 1-2 years of experience in the position) to disseminate knowledge to others.

\begin{myframe}{Observation 5} 
Although all three social attributes were
seen as strongly positive among the respondents, genders differ in opinion on the most important social attribute. While women think that \textit{communicating effectively} is
critical, the men's group thinks that \textit{sharing knowledge with the team} is more important for a great SQA engineer.
\end{myframe}

\subsubsection{\textbf{Personal Attributes}}

\begin{figure*}[!tb]
\centering{\includegraphics[scale=0.6]{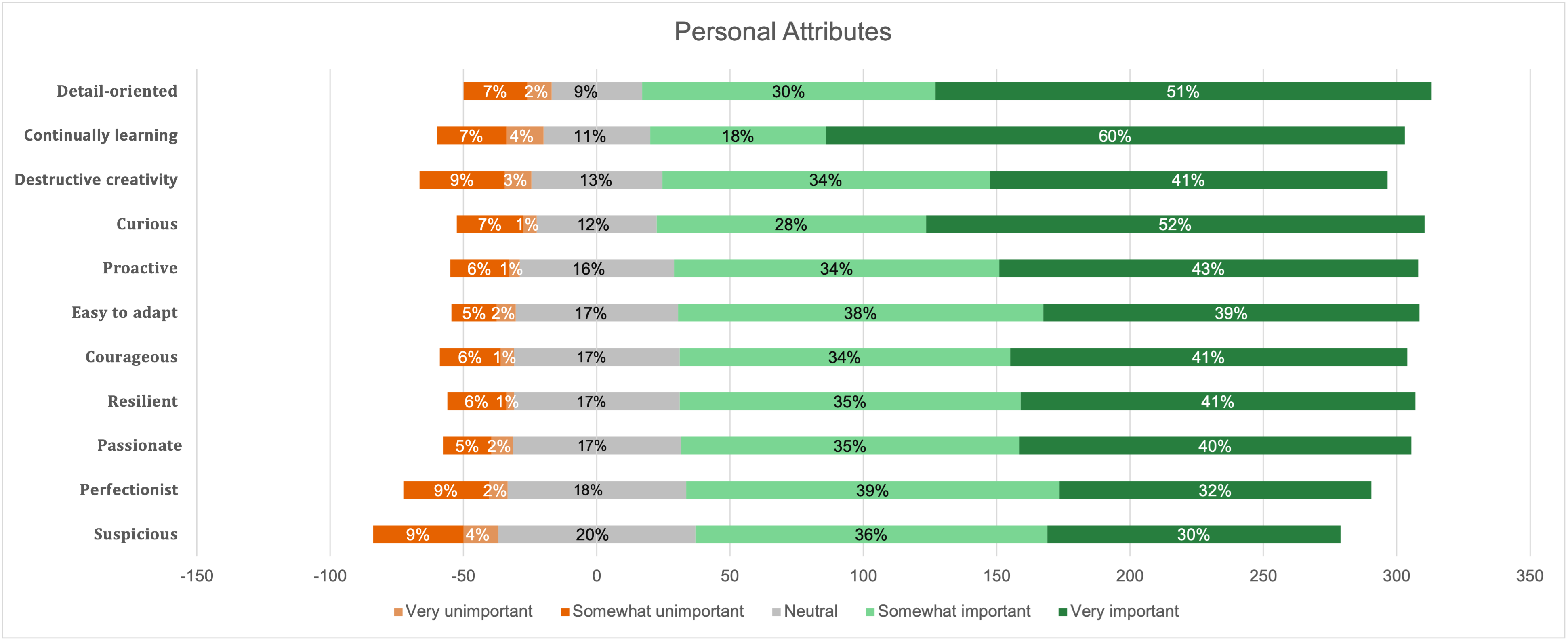}}
\caption{Importance of Personal Attributes reported by the respondents.
}
\label{personal1}
\end{figure*}

\begin{table*}[!tp]
\caption{Results of the rating survey for the personal category per gender}
\label{tab:personalGender}
\centering
\begin{tabular}{@{}lccccc@{}}
\cmidrule(l){2-6} & \multicolumn{5}{c}{\textbf{Female}}                                                     \\ \cmidrule(l){2-6} & Very important & Somewhat important & Neutral & Somewhat unimportant & Very unimportant \\
Continually Learning   & 69.9\%         & 12.0\%             & 9.0\%   & 6.8\%                & 2.3\%            \\
Curious                & 62.4\%         & 21.8\%             & 12.8\%  & 3.0\%                & 0.0\%            \\
Detail Oriented        & 60.2\%         & 26.3\%             & 9.8\%   & 3.8\%                & 0.0\%            \\
Passionate             & 56.4\%         & 29.3\%             & 12.8\%  & 1.5\%                & 0.0\%            \\
Proactive              & 55.6\%         & 32.3\%             & 9.8\%   & 2.3\%                & 0.0\%            \\
Resilient              & 54.1\%         & 28.6\%             & 14.3\%  & 3.0\%                & 0.0\%            \\
Destructive Creativity & 52.6\%         & 27.8\%             & 11.3\%  & 6.0\%                & 2.3\%            \\
Easy to Adapt          & 51.1\%         & 34.6\%             & 11.3\%  & 2.3\%                & 0.8\%            \\
Courageous             & 51.1\%         & 30.8\%             & 13.5\%  & 3.8\%                & 0.8\%            \\
Suspicious             & 39.8\%         & 41.4\%             & 8.3\%   & 6.8\%                & 3.8\%            \\
Perfectionist          & 39.1\%         & 42.1\%             & 13.5\%  & 5.3\%                & 0.0\%            \\ \cmidrule(l){2-6} 
& \multicolumn{5}{c}{\textbf{Male}} \\ \cmidrule(l){2-6} & Very important & Somewhat important & Neutral & Somewhat unimportant & Very unimportant \\
Continually Learning   & 53.5\%         & 21.9\%             & 12.3\%  & 7.5\%                & 4.8\%            \\
Detail Oriented        & 46.1\%         & 32.5\%             & 9.2\%   & 8.3\%                & 3.9\%            \\
Curious                & 45.6\%         & 30.7\%             & 12.3\%  & 9.2\%                & 2.2\%            \\
Proactive              & 36.0\%         & 34.2\%             & 19.7\%  & 8.3\%                & 1.8\%            \\
Courageous             & 35.5\%         & 36.0\%             & 19.3\%  & 7.5\%                & 1.8\%            \\
Destructive Creativity & 34.6\%         & 37.3\%             & 14.5\%  & 10.5\%               & 3.1\%            \\
Resilient              & 33.3\%         & 39.0\%             & 18.9\%  & 7.9\%                & 0.9\%            \\
Easy to Adapt          & 32.0\%         & 39.5\%             & 20.2\%  & 5.7\%                & 2.6\%            \\
Passionate             & 31.6\%         & 38.2\%             & 20.2\%  & 7.0\%                & 3.1\%            \\
Perfectionist          & 28.5\%         & 36.8\%             & 21.1\%  & 11.0\%               & 2.6\%            \\
Suspicious             & 24.6\%         & 33.8\%             & 27.6\%  & 11.0\%               & 3.1\%           
\end{tabular}%
\end{table*}

\begin{table*}[!tb]
\caption{Results of the rating survey for the personal category per experience level}
\label{tab:personalExp}
\centering
\begin{tabular}{@{}lccccc@{}}
\cmidrule(l){2-6}
                       & \multicolumn{5}{c}{\textbf{Less Experienced}}                                           \\ \cmidrule(l){2-6} 
                       & Very important & Somewhat important & Neutral & Somewhat unimportant & Very unimportant \\
Continually Learning   & 77.9\%         & 10.3\%             & 4.4\%   & 1.5\%                & 5.9\%            \\
Curious                & 63.2\%         & 25.0\%             & 7.4\%   & 2.9\%                & 1.5\%            \\
Detail Oriented        & 58.8\%         & 27.9\%             & 7.4\%   & 2.9\%                & 2.9\%            \\
Resilient              & 50.0\%         & 30.9\%             & 14.7\%  & 2.9\%                & 1.5\%            \\
Courageous             & 48.5\%         & 29.4\%             & 13.2\%  & 5.9\%                & 2.9\%            \\
Proactive              & 47.1\%         & 33.8\%             & 14.7\%  & 4.4\%                & 0.0\%            \\
Passionate             & 45.6\%         & 32.4\%             & 17.6\%  & 1.5\%                & 2.9\%            \\
Easy to Adapt          & 44.1\%         & 38.2\%             & 13.2\%  & 2.9\%                & 1.5\%            \\
Destructive Creativity & 36.8\%         & 41.2\%             & 7.4\%   & 8.8\%                & 5.9\%            \\
Suspicious             & 30.9\%         & 39.7\%             & 16.2\%  & 5.9\%                & 7.4\%            \\
Perfectionist          & 23.5\%         & 47.1\%             & 20.6\%  & 7.4\%                & 1.5\%            \\ \cmidrule(l){2-6} 
                       & \multicolumn{5}{c}{\textbf{Experienced}}                                                \\ \cmidrule(l){2-6} 
                       & Very important & Somewhat important & Neutral & Somewhat unimportant & Very unimportant \\
Continually Learning   & 56.4\%         & 18.0\%             & 11.0\%  & 10.5\%               & 4.1\%            \\
Detail Oriented        & 53.5\%         & 26.7\%             & 9.9\%   & 7.6\%                & 2.3\%            \\
Curious                & 51.2\%         & 27.9\%             & 12.2\%  & 8.1\%                & 0.6\%            \\
Destructive Creativity & 44.2\%         & 27.3\%             & 17.4\%  & 9.9\%                & 1.2\%            \\
Proactive              & 43.0\%         & 30.8\%             & 18.0\%  & 7.0\%                & 1.2\%            \\
Resilient              & 40.7\%         & 32.6\%             & 19.2\%  & 7.6\%                & 0.0\%            \\
Perfectionist          & 40.1\%         & 32.0\%             & 15.7\%  & 9.9\%                & 2.3\%            \\
Passionate             & 38.4\%         & 34.3\%             & 19.8\%  & 6.4\%                & 1.2\%            \\
Courageous             & 36.0\%         & 34.3\%             & 20.9\%  & 7.6\%                & 1.2\%            \\
Suspicious             & 36.0\%         & 30.2\%             & 21.5\%  & 10.5\%               & 1.7\%            \\
Easy to Adapt          & 34.9\%         & 39.0\%             & 19.8\%  & 5.2\%                & 1.2\%            \\ \cmidrule(l){2-6} 
                       & \multicolumn{5}{c}{\textbf{Very Experienced}}                                           \\ \cmidrule(l){2-6} 
                       & Very important & Somewhat important & Neutral & Somewhat unimportant & Very unimportant \\
Continually Learning   & 54.5\%         & 22.8\%             & 14.6\%  & 5.7\%                & 2.4\%            \\
Curious                & 45.5\%         & 29.3\%             & 15.4\%  & 7.3\%                & 2.4\%            \\
Courageous             & 43.9\%         & 36.6\%             & 13.8\%  & 4.9\%                & 0.8\%            \\
Detail Oriented        & 43.9\%         & 36.6\%             & 9.8\%   & 7.3\%                & 2.4\%            \\
Proactive              & 41.5\%         & 37.4\%             & 13.8\%  & 5.7\%                & 1.6\%            \\
Easy to Adapt          & 41.5\%         & 35.8\%             & 14.6\%  & 4.9\%                & 3.3\%            \\
Passionate             & 40.7\%         & 37.4\%             & 13.8\%  & 4.9\%                & 3.3\%            \\
Destructive Creativity & 39.0\%         & 39.0\%             & 11.4\%  & 7.3\%                & 3.3\%            \\
Resilient              & 35.8\%         & 41.5\%             & 15.4\%  & 5.7\%                & 1.6\%            \\
Perfectionist          & 26.0\%         & 43.1\%             & 21.1\%  & 8.1\%                & 1.6\%            \\
Suspicious             & 22.0\%         & 43.1\%             & 21.1\%  & 9.8\%                & 4.1\%           
\end{tabular}%
\end{table*}

Among all the eleven attributes described in the personal category, we can highlight the following top three personal attributes according to the participants: \textit{continually learning}, which refers to building new skills, seeking new improvements, and continuously searching for the knowledge, received 60\% of the votes as ``very important". This aligns with our findings from the interviews since most of the participants stated that quick learning ability is very important to becoming a great SQA engineer.
\textit{Curious}, which refers to investigating beyond what has been specified, and questioning everything until fully understanding what is going to be tested, received 52\% of the votes as ``very important". Followed by being \textit{detail-oriented}, which refers to paying attention to details when testing a feature or when reporting a bug. Fifty-one percent of the participants think that it is very important to write like a sophisticated journalist when creating a test script or a test plan. The entire list of personal attributes is listed in Figure \ref{personal1}.

\textbf{Contextual factor - Gender}\\
Because this is a large category, it is interesting to look at different contextual factors to investigate differences that we could not observe when looking into the entire population. Although the top three attributes remained the same among the genders, we can notice some differences in the order of priority (see Table \ref{tab:personalGender}). For instance, the attribute \textit{passionate}, which is being intrinsically interested in the area they work on, seems to be more valued among the female group while the attribute \textit{courageous}, which is being willing to say ``no" to allowing things to be released when they are not comfortable approving it, seems to be more valued between the male group. 

\textbf{Contextual factor - Experience}\\
When we split the survey data for the personal category across levels of experience (see Table \ref{tab:personalExp}), we noticed that some attributes started to gain more importance for more experienced SQA engineers. For example, the attributes \textit{destructive creatively}, which refers to finding creative ways to break the system, and the attribute \textit{easy to adapt}, which refers to handling changes in priorities well, adapting to the company's context. Also, we noticed that some attributes started to lose some importance for more experienced SQAs. For example, the attribute \textit{resilience}, which is the ability to withstand or recover quickly from difficult conditions, had 50\% votes as ``very important" among the less experienced SQAs but only 35.8\% for the very experienced ones. 

\begin{myframe}{Observation 6} 
The top three personal
attributes are \textit{continually learning},
being \textit{curious}, and \textit{detail-oriented}.
\end{myframe}

\subsection{RQ3 - How do the attributes of a great SQA engineer differ from other software development roles?} 
\label{results3}

Previous work identified several attributes for great software engineers (SE) \cite{Li:2015,Li:2020}, great managers \cite{Kalliamvakou:2019}, and great Open Source Software (OSS) maintainers \cite{dias:2021}. As these attributes are specific to the roles (software engineers, managers, and OSS maintainers), some attributes seem more valuable to one role than others. For example, the attribute \textit{enables autonomy} by providing freedom on how engineers work was only mentioned for managers, and the attribute \textit{transparency} by making available information about the project was only mentioned for OSS maintainers.

Table \ref{tab:comparison} shows the attributes of a great SQA engineer in comparison with these previous studies. We found that the important attributes of SQA and SE are very similar. Twenty-one out of twenty-five attributes of a great SQA were mentioned in the studies with software engineers. It can be expected since they are usually part of the same team and the work context and environments for SQA and SE are much the same.

We also found that five attributes of great SQAs were already uncovered in all three related studies, most of the time with different wording but having the same meaning. For instance, the attribute of \textit{being a leader without the title} was mentioned in the great SE studies as \textit{walking-the-walk and mentoring}, a SE that acts as the exemplar for others to follow; while it was mentioned as \textit{guide the team} and \textit{inspire the team} for managers and as \textit{leadership} for OSS contributors. 

\textit{Communication} was key for all studies. It is mentioned in the great SE studies as the SE who creates a shared understanding with others; a great manager mediates communication, and as communication for OSS contributors as to have the ability to exchange information in a sensible way.

Regarding the attribute \textit{collaborates with others}, the studies about great SE brought a similar characteristic of ``creating share context"; while the study about great managers named it as ``clearing a path to execution", and the study about OSS contributors brought it as ``community building". All studies used different terms with the same meaning of helping encourage new team members to keep participating in the project.

\textit{Technical expertise} for SE refers to being knowledgeable about tools and building materials, while it is \textit{being technical} for managers and \textit{having technical excellence} for OSS contributors.

\begin{myframe}{Observation 7} 
\textit{Being a leader without the title, communicating effectively, collaborating with others, technical expertise, and sharing knowledge with others} had different wording in previous studies but are considered very important attributes for other software development roles. 
\end{myframe}

Nonetheless, four attributes were first identified in our study as they are terms more likely used for the quality assurance role. For instance, being \textit{courageous} to say no only makes sense to SQA engineers since they are responsible for approving or disapproving a feature, software, or release code after testing it. Similarly, having \textit{knowledge about the QA area} is required for SQA engineers but not for other software development roles. Although the attributes \textit{
dealing well with priorities} and being \textit{perfectionist} could be applied to the other roles, they were first identified in our study. A possible reason is that an SQA engineer usually gets multiple features to test at the same time and has to stop testing something due to failures, or problems in the code, and start testing something else that is available at the backlog. Also, a possible reason for perfectionism being more common for SQAs is that they need to guarantee the quality of the deliverable.				

\begin{myframe}{Observation 8}
The attributes that were only identified in this study with SQA engineers are: being \textit{courageous, having knowledge about the QA area, dealing well with priorities, and being perfectionist}. 
\end{myframe}

\begin{table}[!tbp]
\caption{Comparison of attributes of a great SQA with related work}
\label{tab:comparison}
\resizebox{\columnwidth}{!}{%
\begin{tabular}{@{}lcccc@{}}
\toprule
Attribute &
  \begin{tabular}[c]{@{}c@{}}Great SQA\\ (our study)\end{tabular} &
  \begin{tabular}[c]{@{}c@{}}Great SE\\ (2015 \& 2019)\end{tabular} &
  \begin{tabular}[c]{@{}c@{}}Great Manager\\ (2019)\end{tabular} &
  \begin{tabular}[c]{@{}c@{}}Great OSS \\ contributors\\ (2021)\end{tabular} \\ \midrule
Being a leader without the title & x & x                    & x                    & x                    \\
Communicates effectively         & x & x                    & x                    & x                    \\
Collaborates with others         & x & x                    & x                    & x                    \\
Technical expertise              & x & x                    & x                    & x                    \\
Shares knowledge with others     & x & x                    & x                    & x                    \\
Continuous learning              & x & x                    &                      & x                    \\
Programming skills               & x & x                    &                      & x                    \\
Knowledgeable about domain     & x & x                    &                      & x                    \\
Analytical and systematical view & x & x                    &                      &                      \\
Detail-oriented                  & x & x                    &                      &                      \\
Destructive creativity           & x & x                    &                      &                      \\
Curious                          & x & x                    &                      &                      \\
Proactive                        & x & x                    &                      &                      \\
Resilience                       & x & x                    &                      &                      \\
Easy to adapt                    & x & x                    &                      &                      \\
Passionate                       & x & x                    &                      &                      \\
Holistic view                    & x & x                    &                      &                      \\
Customer-centric                 & x & x                    & \multicolumn{1}{l}{} & \multicolumn{1}{l}{} \\
Suspicious                       & x & x                    & \multicolumn{1}{l}{} & \multicolumn{1}{l}{} \\
Think critically                 & x & x                    & \multicolumn{1}{l}{} & \multicolumn{1}{l}{} \\
Time management                  & x & x                    & \multicolumn{1}{l}{} & \multicolumn{1}{l}{} \\
Courageous                       & x & \multicolumn{1}{l}{} & \multicolumn{1}{l}{} & \multicolumn{1}{l}{} \\
Dealing well with priorities     & x & \multicolumn{1}{l}{} & \multicolumn{1}{l}{} & \multicolumn{1}{l}{} \\
Perfectionist                    & x & \multicolumn{1}{l}{} & \multicolumn{1}{l}{} & \multicolumn{1}{l}{} \\
Knowledge about the QA area      & x &                      &                      &                    
\end{tabular}
}
\end{table}

\section{Discussion}
\label{discussion}

Although most of the attributes of great software quality assurance (SQA) engineers that we identified have been previously discovered in prior work (Section \ref{results3}), our findings are the first study that brings a comprehensive set of attributes for SQAs and compare with related work. This section presents in more detail aspects uncovered with this work and its implications for practitioners, researchers, and educators. 

\subsection{Implications for practitioners}
\label{rec}

From analyzing the data from the interviews and survey, we highlight some observations for SQA practitioners as ways of being great at their jobs that emerged with important implications.

According to our observation 6, \textit{continuous learning} is in the top three of the personal attributes. It was considered a very important attribute for all groups explored in this study: general, gender, and experience. As the SQA becomes more experienced, they keep thinking that this is very important:
    
   \textit{``We need to look for new improvements because the IT market changes every day".} - \#P5, Senior SQA Engineer.

This finding confirms a previous work with software engineers \cite{Li:2020} where informants felt that engineers with higher levels of experience understood that they needed to be continuously improving to stay ahead.

Some participants of our study stated that a good way to become an expert in a subject and learn continuously is to produce technical content about a subject and present it to a group of people. Practitioners may introduce it at work as a way of becoming experts in something. Our findings suggest that being a continuous learner and having multidisciplinary skills help the SQA professional exchange positions easier if they desire or if the market changes in a different direction.

Regarding technical attributes, our observation 4 showed that \textit{having theoretical knowledge of the SQA area} was considered the most important attribute of the technical category regardless of the contextual factors of gender and experience. Our results suggest that being able to learn the theory of the QA area and having technical expertise is likely more important than having programming skills. Practitioners may think it interesting to take testing certifications such as the International Software Testing Qualifications Board (ISTQB) \footnote{https://www.istqb.org/} among others as well as considering these findings when hiring new employees.

The third observation highlighted the decision-making attributes that were considered very important by the respondents. Informants stated that being customer-centric helps to make effective decisions and perform better testing.

\textit{``When we think as a final user, we go beyond what was specified"} (\#P8). 
A great SQA engineer needs to become more customer-centric by providing generous customer service, anticipating customer needs, collecting customer feedback, and embodying its values and company culture. 

Finally, \textit{collaborating with others} and \textit{sharing knowledge with the team} are examples of social attributes of a great SQA engineer. During the interviews, the participants were asked about the relationship between developers and SQAs (testers) and they stated that there is still some resistance among them (\#P13) but the relationship between developers and SQAs tends to improve when both understand the definition of ``done". The functionality should not be done when the developer finishes coding or when the SQA completes testing it, but when they work together as a team until the functionality meets all the acceptance criteria.
There are several ways to collaborate with the team. For instance, by implementing pair programming where the SQA is the co-pilot, developers help with test automation/write test code, SQAs review development code, and so on. Also, our results indicate that when the company adopts the same programming language for development and testing, the developer and SQA collaboration tends to increase since they can understand and support each other's code.

\subsection{Implications for researchers}

We considered two contextual factors in this work: gender and experience. It allowed us to perceive implications and research opportunities that we would not be able to notice when considering the whole group of respondents.

Considering the decision-making attributes grouped by gender, our results suggest that women think that being a customer-centric SQA is more important, while men consider \textit{taking an analytical and systematical view} more important. Future work could look at gender for other attributes to see if there is any correlation and/or possible reason for this. 

Still looking at gender but for the \textit{social attributes}, our study shows that genders differ in opinion on the most important attribute. 69\% of women think that \textit{communicating effectively} is very important for a great SQA engineer while only 44.7\% of men think that. Future research may explore how women SQA communicates at work and why they think it is very important for this role.

Regarding the personal attributes of a great SQA, only a few attributes in this category seemed to make a difference between the genders. Our results reveal that women think of being \textit{passionate} for their jobs more than men. On the other hand, the men group expressed that being \textit{courageous} is more valued than passionate. 

From our observation 2, \textit{time management} and \textit{dealing well with priorities} were considered critical by the respondents, and as
the SQA becomes more experienced, \textit{time management}
becomes more critical and gains more importance. Since it seems to be a common thing among software engineers and SQAs (see Table \ref{tab:comparison}), this finding raises a research opportunity to compare the existing tools in the market used to manage work time and develop one specific to meet developer and SQAs needs. Similarly, future work could evaluate the most effective tools to help manage priorities and highlight practical recommendations to the industry.

Our third research question explored how the attributes of a great SQA engineer differ from other attributes identified in the previous studies (software engineers, project managers, and maintainers). We noticed that each study named attributes differently with descriptions that mean nearly the same as another word or phrase. After we compared and merged similar topics, we found that 84\% of the attributes of great SQAs are in common with software engineers (see Table \ref{tab:comparison}). Future studies could examine these attributes to understand why they are important for both roles and which scenario they apply to.

\subsection{Implications for educators}

After identifying that \textit{having theoretical knowledge of the SQA area} is the most important attribute of the technical category (observation 4), and hearing from participants who learned about it in practice, it brought us the following concern: what has been covered on software quality assurance courses?
The curriculum of software engineering varies from place to place but to the best of our knowledge, there is no course to learn about the fundamentals of testing or the career paths for developers/testers. Educators may think of adding it to the Software Engineering curriculum aiming to prepare great professionals for the market.

\section{Threats to Validity}
\label{threats}

In this section, we discuss some potential threats to the validity of our study with their corresponding mitigating factors according to \cite{Runeson:2008}, which distinguish the following criteria for validity:

\subsection{Construct Validity}
Construct validity concerns the extent to which our study accurately assesses what it's supposed to. For this study, we performed two empirical methods: interview and survey. For each method, we mitigated risks by conducting pilot studies to adjust the time required to perform both the interview and survey, the quality of our questions, and thus mitigate any construct threats. For the survey questions, we made sure the questions were clear by adding the meaning of each attribute and/or examples to it.

Another concern refers to the saturation of the results. To mitigate this threat, we felt that reached a comprehensive set of attributes of great SQA engineers until no new attributes stopped emerging during new interviews.

Finally, we guaranteed the anonymity of the participants and all information gathered during the interviews and survey
will be used only by the research team.

\subsection{Internal Validity}
Misinterpretation of the interviews is an internal validity issue we need to consider.  Moreover, the data analysis of both studies could have been biased by the researcher's opinion. To mitigate these threats, each step was discussed among the authors, refined, brought to a consensus through mutual agreement, and merged results to shape the final result. 
We also used the coding technique to analyze the interview transcripts, comparing our findings with previous ones.
Regarding the survey, some of the questions might have been misunderstood by the participants. The pilot studies were critical to assess whether the questions were unambiguous.

\subsection{External Validity}
The main threat to external validity is the non-generalizability of the study. We did not sample professionals in the Software Engineering field besides SQA engineers, but we believe that by selecting SQA professionals from different companies around the world we were able to vary by experience level, methodology, objective, and culture. For the interviews, we selected participants from Brazil, USA, India, and Portugal (a total of 25 practitioners), while the survey's respondents were from 18 different nationalities spread across all continents. 

We also performed a large survey to validate the results from the interviews with a large number of participants. We received many responses from females (133 responses, 36.6\%), which allowed us to highlight systematic gender differences that in more in-depth studies of interesting sub-populations are worthwhile.
Thus, the profiles of both studies were diverse, and the majority had more than five years of experience as SQA, which we believe helped mitigate the threat of the non-generalizability of the study. 

\section{Conclusion}
\label{conclusion}

In this paper, we presented an investigation of the software quality assurance role and highlighted which attributes were perceived as most important by them. We also performed a survey with a large number of SQA practitioners around the world to assess the importance of each of the identified attributes from the interviews and investigate contextual aspects (gender and experience) of their opinions. 

Although the daily responsibility of an SQA engineer is to test the software and guarantee the quality of the entire software development cycle, this study showed that a great SQA has to overpower a lot of soft skills, and social, technical, management, and decision-making attributes. 

The top five distinguishing characteristics of a great SQA engineer are curiosity, ability to communicate effectively, critical thinking skills, learning continuously, and technical expertise. On that account, we built evidence that showed that SQAs have a vital role in any organization, and they act as a bridge among all the parties that work to deliver a product.  
For future work, we intend to perform more empirical studies exploring more contexts (e.g. age, personality, demography information), triangulating the way forward toward deepening our
understanding of what makes great software quality assurance engineers. Thinking more broadly and taking advantage of the neuroscience area in software engineering, we want to take the main characteristics as insights into a brain study with SQA engineers to collect their biofeedback through brain behaviors to distinguish great SQA engineers from ordinary ones.

\appendices


\ifCLASSOPTIONcompsoc
  \section*{Acknowledgments}
\else
  \section*{Acknowledgment}
\fi

The authors would like to thank all the participants for agreeing to participate in this study, either through interviews or by answering the survey.

\ifCLASSOPTIONcaptionsoff
  \newpage
\fi


\bibliographystyle{IEEEtran}
\bibliography{bib}

\end{document}